\documentclass[twocolumn,showpacs,preprintnumbers,amsmath,amssymb]{revtex4}    
\usepackage{graphicx}
\usepackage{dcolumn}
\usepackage{bm}
\begin{document}  
\title{Light Localization and Lasing in a 3D Random Array of Small Particles}  
 \author{  
Paul.~R.~Sievert }  
\affiliation{ 	  
Condensed Matter/ Electronic Structure Theory Group,   
Department of Physics and Astronomy,
Northwestern University, Evanston, IL  
60208-3112. \\ } 
\date{\today}  
\begin{abstract}
The results of computer simulations of light scattering by a random array of small particles is presented.
A tensor Green's function method is employed. Results are
given for arrays of particles situated randomly in a cubic, 3D sample $1.6\mu$ on a side with particle numbers ranging from
100 to 937. The material parameters used correspond to ZnO and Ag, over the 
wavelength range $300nm < \lambda < 400nm$. The particle diameter considered for ZnO is 50nm  and for
Ag 20nm. The eigenvalue spectrum of the scattering matrix is presented and the magnitude of the lowest eigenvalue
is used to estimate the extent of light localization. The scattering problem was solved for a unit amplitude incident
plane wave. The energy densities and losses, both total and scattering, were calculated.
No evidence of light localization was found, neither in the existence of a zero eigenvalue nor in any low lying
isolated resonance, although the lowest eigenvalues for Ag are an order of magnitude smaller than those for ZnO. Energy density
calculations show that energy is mostly confined to the region near and in the scattering particles, and that 
little energy concentrates
in the fields between particles.
The energy-energy correlation function shows a peak at the mean particle separation in a universal manner.
Mixtures of Ag and ZnO particles show isolated spectra with little interaction consistent with this 
energy localization. 
Gain was added to the ZnO system by adding a resonant term having Lorentzian line shape with a negative coefficient, which
drives the imaginary part of the dielectric constant negative corresponding to gain. With gain in the system, the
losses can be compensated; however, none of the eigenvalues moves toward zero, indicating that localization and 
loss compensation (which leads to a lasing threshold) are not connected. These results indicate that similar calculations with
a larger number of scattering sites would still be of interest. This, however, will
require very different numerical techniques than have hitherto been applied.     
\end{abstract}
\pagebreak

\maketitle
\section{Introduction}
Light localization due to coherent multiple scattering and interference in random systems has been
the subject of much speculation and investigation since the late 1990's. This speculation arises in connection 
with experimental studies of laser action in random media by Lawandy et al.\cite{Lawandy}. Their results indicated that an
earlier diffusion transport theory of the lasing phenomenon by Letokhov\cite{Letokhov} was inadequate to explain
the low threshold activation energy for laser action. Light localization due to interference
was suggested by Wiersma et al.\cite{Wiersma}. Experimental evidence was presented, although more recent work has called this into
question and indicates an important roll for absorption\cite{ChabStoy}. Also, this later experimental work, using
microwaves, seems to rule out localization
in 3D.  Localization was suggested in an   
analogy with electrons in 3D random systems which has been shown will localize\cite{Anderson}\cite{John}.
The surmised localization condition for light is that the mean free path due to scattering $l_{scatt}$ be approximately equal to
$\lambda/2\pi$ and that the scattering be ``strong''. This is the Ioffe-Regel\cite{Ioffe} condition for localization with a random array
of scatterers in 3D. One should be able to check this with a computer simulation. An unambiguous criterion for a local state
to exist is the vanishing or near vanishing of any of the eigenvalues of the scattering operator. Near localization would
manifest itself as a sharp separated resonance near zero. Trends in the variance of the eigenvalue spectrum with the number
of scattering sites and frequency could also measure the extent and onset of localization\cite{Chabanov}.
Earlier simulations of light in 3D scattering\cite{Poles} have used no more than 100 scattering sites. The eigenvalue spectra of 
these systems did not show any appreciable localization.
Since the absence of localization may be a consequence of system size, we are motivated to 
repeat these attempts with many more scattering sites and to examine the scattering operator
eigenvalue spectrum and also the energy density distribution when incident
light is present; this should give some picture of the extent of such localization.

Accordingly, we present results here for systems with up to 900 scattering sites which seems to be near the 
upper limit for direct numerical methods that still yield convergent results. In this paper, we use material parameters
appropriate to ZnO for dielectrics and for Ag for metals and confine ourselves to the $300nm \leq \lambda \leq 400nm$ wavelength
range.
These particle sizes are consistent with the long wavelength limit.
  
Further, there have been reports of induced lasing 
in samples of random ZnO particles in a powder\cite{CaoXu}\cite{Cao} and a calculation of a lasing threshold
in a model with up to 900 scattering sites\cite{Burin}. Light localization has been cited as a mechanism that would
provide the feedback needed for lasing action in these 3D systems. The connection of localization and lasing threshold
for 1D systems seems to be well established\cite{Qiming} but it has been questioned for 3D systems\cite{Soest}\cite{Vanneste}.   
Hence an investigation of the connection of this lasing threshold with localization
is therefore warranted.

The theoretical formulation of the light scattering problem is presented in Sect. II, whose details are relegated to an Appendix;
together these give a review of the vector field scattering theory employed here. Its generalizations are readily apparent and will be 
published elsewhere.
The  material model to which this theory is applied is presented in Sect. III along with the formulation of quantities for which numerical
results are presented. Eigenvalue and scattering results are described in Sect. IV. In Sect. V nonradiative gain
is considered. The conclusions which can be gleaned from this work are stated
in Sect. VI.

\section{Theoretical Formulation}

The clearest formulation of the light scattering problem in random media is
probably given by a 
multiple scattering method using a free particle propagator or Green's 
function appropriate to photons and scattering centers, randomly situated,
that represent the medium. This formulation is derived in a very general fashion in the
Appendix.

The solution for the electric field has the scattering equation form:
\begin{equation}
\vec{E}\left(\vec{r}\right) = \vec{E}_{o}\left(\vec{r}\right) +
\int d^{3}r^{\hskip .5mm '}\stackrel{\leftrightarrow}{G}(\vec{r}-\vec{r}^{\hskip 1mm '})
\cdot k^{2}4 \pi \vec{P}(\vec{r}{\hskip .5mm '})
\label{Scatt}
\end{equation}
where $\vec{r}^{\hskip 1mm '}$ is the 
source location, $\vec{r}$ the observation position 
and we define a tensor Green's function as
\begin {equation}
\stackrel{\leftrightarrow}{G}(\vec{r}-\vec{r}^{\hskip 1mm '})=\left ( \stackrel{\leftrightarrow}{I}+
\frac{ \vec{\nabla}\vec{\nabla}}{ k^2}\right ) g(|\vec{r}-\vec{r}^{\hskip 1mm '}|).
\label{Dyadic}
\end{equation}
Expanded, Eq.(\ref{Dyadic}) becomes
\begin{widetext}
\begin{equation}   
\stackrel{\leftrightarrow}{G}(\vec{R})=
\left[ \stackrel{\leftrightarrow}{I} \left ( 
1+\left(\frac{ikR-1}{k^2 R^2} \right) \right)         
+ \left( \frac{3-3ikR-k^2 R^2}{k^2 R^4}\right)\vec{R}\vec{R}
\right]g\left(R\right)
\label{ExpG}           
\end{equation}
\end{widetext}
where
\begin{equation}
g\left(R\right) =  \frac{e^{ikR}}{4\pi R}
\end{equation}
is the scalar retarded Green's function.
In the usual continuum formulation of the light scattering problem \cite{Martin}, the polarizability
is identified as
\begin{equation}
4\pi\vec{P}\left(r\right) \equiv \Delta\epsilon\left(\vec{r}\right)\vec{E}\left(\vec{r}\right)
\label{Polar}      
\end{equation}
and
\begin{equation}
\Delta\epsilon\left(\vec{r}\right) = \left(\epsilon\left(\vec{r}\right)-\epsilon_{o}\right)
\label{Epsilon}.
\end{equation}
Here $\epsilon\left(\vec{r}\right)$ is the position dependent dielectric constant and $\epsilon_{o} = 1$ in Gaussian units.
We assume here linear response. Other, higher order, induced moments may not respond as simply, but we
will stick to dipoles as the elementary sources.
It should be pointed out here that in the continuum formulation, where a spatial varying
dielectric function is used, the dielectric function could have a general tensor
form. Other boundary conditions can be easily applied. For example, there are
many formulations for implementing periodic boundary conditions for the scalar Green's
function \cite{Periodic} and carrying this over to the tensor case is quite easy as
the tensor Green's function is obtained by local operations, i.e. derivatives of the
scalar Green's function, $g\left(R\right)$.
\begin{figure*}
\includegraphics[width=6in,angle=0]{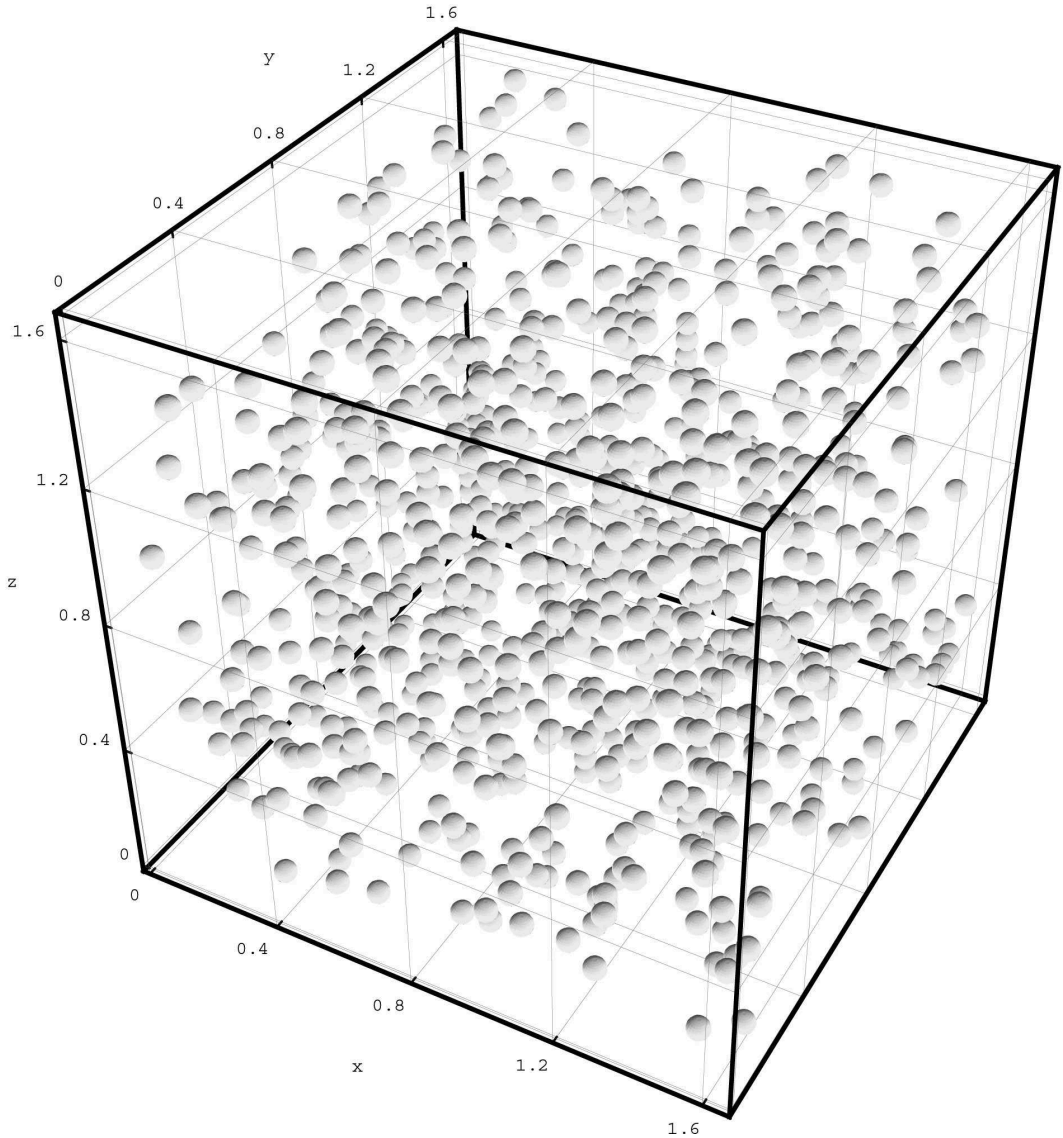}
\caption{\label{fig:50nmZnO900} The model sample $1.6\mu$ on a side showing 900 randomly distributed 50nm diameter ZnO spheres.
In this case the grid on which they are distributed is cubic with 50nm spacing.}
\end{figure*}

Here we consider a random medium consisting of N discrete grains. We shall take the typical dimension of all grains
to be $a_i \ll \lambda$, where  $\lambda$ is the light wavelength and $i:1\rightarrow N $.  The integral in Eq.(\ref{Scatt})
now becomes a sum of integrals, one over each grain. We use $\vec{r}_i$ and $\vec{r}_j$ to label the 
location of the centers of the discrete grains i and j, respectively. For all integrals over grains where
$\vec{r}_i^{\hskip .5mm '}\neq\vec{r}_j$, Eq.(\ref{Polar}) can be used and the $\vec{E}\left(\vec{r}_j\right)$
taken as constant over the grain as is appropriate in the long wavelength limit. For the case where
$\vec{r}_i^{\hskip .5mm '} = \vec{r}_j$, the integral over this ``diagonal grain'' is singular and the integral
must be considered as a principal value integral. The singularity is extracted as
\begin{equation}
\stackrel{\leftrightarrow}{G}(\vec{r}-\vec{r}^{\hskip 1mm '})= P.V.
\stackrel{\leftrightarrow}{G}(\vec{r}-\vec{r}^{\hskip 1mm '})-\frac{\stackrel{\leftrightarrow}{L}
\delta (\vec{r}-\vec{r}^{\hskip 1mm '})}{k} 
\end{equation}
where $\stackrel{\leftrightarrow}{L}$ is due to an exclusion volume at the center of the diagonal grain.
This singularity is due to taking both $\vec{\nabla}$'s in the definition of Eq.(\ref{Dyadic}) under the
integral sign in the vector scattering equation, Eq.(\ref{Scatt}). Mathematically this exclusion or 
depolarization volume shape
depends on the shape of the grain and must be needle shaped if the grain is so shaped etc. This shape
dependence follows from a choice in the order of taking limits in defining the principle value.
These considerations are warranted due to the scale invariance of Laplace's equation. This singularity has been extensively
studied by many authors \cite{Chew},\cite{Yaghjian} and its extraction is central to the derivation
of the Lorentz-Lorenz equation when going to the continuum limit inside a
scattering medium \cite{Born}. Such mathematical considerations seem non-physical and the proper limiting proceedure
 may be application dependent. Lorentz originally assumed spherical or cubic symmetry\cite{Lorentz}, although he indicated the possible existence of other terms. In what follows, assume
a spherical diagonal grain; 
with spherical symmetry, $\stackrel{\leftrightarrow}{L}=\frac{\stackrel{\leftrightarrow}{I}}{3}$.
Extracting the singularity and integrating over the diagonal grain, Eq.(\ref{Scatt}) now takes the form
\begin{widetext}
\begin{equation}
\vec{E}\left(\vec{r}_{i}\right) = \left(\frac{1}{D_{i}}\right)\left[\vec{E}_{o}\left(\vec{r}_{i}\right) +
\sum_{\vec{r}_{j}\neq\vec{r}_{i}}^{N}v_{g_{j}}\stackrel{\leftrightarrow}{G}\left(\vec{r}_{i},\vec{r}_{j}\right)
\cdot k^{2}\Delta\epsilon\left(\vec{r}_{j}\right)\vec{E}\left(\vec{r}_{j}\right)\right]
\label{DisScatt}
\end{equation}
\end{widetext}  
where $v_{g_{j}}$ is the volume of the grain at site $j$, and $D_{i}$, the diagonal term for site $i$, 
given by
\begin{equation}  
D_{i} = 1 - \Delta\epsilon\left(\vec{r}_{i}\right)\left[\frac{2}{3}\left(1-ika_i\right)
exp\left(ika_i\right)-1\right]
\label{Diag}
\end{equation} 
where $a_i$ is now the radius of the spherical diagonal grain. This result has been obtained before
\cite{Martin}; however, its significance is further elucidated by expanding to third order in $ka_i$.
Keeping terms to that order and keeping only the first term in the rhs of Eq.(\ref{DisScatt}), we have for the
field inside of a grain:
\begin{equation} 
\vec{E}\left(\vec{r}\right) =  \frac{3\vec{E}_{o}\left(\vec{r}\right)}{\left(\epsilon+2\right) - \left(\epsilon-1\right)
\left(ka\right)^2 - \left(\epsilon-1\right)i\frac{2}{3}\left(ka\right)^3}
\label{CorrField}
\end{equation}
where we have used Eq.(\ref{Epsilon}) and dropped the site index. Equation(\ref{CorrField}) is now recognizable as the Lorentz correction
factor to the incident field for the field inside a grain, but, with additional dynamic terms.
The terms proportional to $k^2$ and $k^3$ are called
the dynamic depolarization
and radiation-damping correction \cite{Meir,Wokaun}, respectively.
If the diagonal grain is left out and we set $D_i = 1$, this Green's function method becomes identical
with what is known as the coupled dipole method \cite{Schatz} in which 
various larger grain geometries are represented by clumping the
individual dipoles. Then, a very large matrix linear scattering problem must be solved with its
attendant convergence problems, especially if metal grains are modeled, in which case, one must
account for plasmons and other dynamic effects due to source kinetic energy.  
The coupled dipole method has been used to check the local field approximation
\begin{equation}
\vec{E}_{loc} = \vec{E} + \frac{4\pi}{3}\vec{P}
\end{equation}
used in deriving the Lorentz-Lorenz formula \cite{Druger} indicating that,
on the average, it is numerically feasible to recover the continuum results
from the coupled dipole method.

Including the diagonal grain allows us to extend the long wavelength approximation
to larger particle sizes \cite{Schatz} and to various shapes, although here we 
restrict ourselves to spheres.	   
The scattering equation, Eq.(\ref{DisScatt}), can be written as
\begin{equation}
\sum_{j=1}^{N}\stackrel{\leftrightarrow}{{\cal P}}\left(\vec{r}_{i},\vec{r}_{j}\right)
\cdot\vec{E}\left(\vec{r}_{j}\right)
 = \vec{E}_{o}\left(\vec{r}_{i}\right),
\end{equation}
where
\begin{widetext} 
\begin{equation}
\stackrel{\leftrightarrow}{{\cal P}}\left(\vec{r}_{i},\vec{r}_{j}\right) = 
\left(\stackrel{\leftrightarrow}{I}D_{i}\delta_{i,j} - 
\left(1 - \delta_{i,j}\right)
\stackrel{\leftrightarrow}{G}\left(\vec{r}_{i}-\vec{r}_{j}\right)\cdot k^{2}
\Delta\epsilon\left(\vec{r}_{j}\right)v_{g_{j}}\right).
\label{bigPolar}
\end{equation}
\end{widetext}
For all scatterers of identical species, the matrix ${\cal P}$ is symmetric in all of its indices or, if species vary, it can
be made symmetric by multiplying both sides by $\Delta\epsilon\left(\vec{r}_{i}\right)v_{g_{i}}$.
However, because the retarded potentials were used, it is not Hermitian. 
Inversion of the complex matrix ${\cal P}$,
solves the scattering problem. Diagonalization of ${\cal P}$ yields complex eigenvalues.
The scattered field $\vec{E}$ is decoupled from the external  field $\vec{E}_o$  
in any mode for which the magnitude of the eigenvalue $\rightarrow 0$. Such a mode can then be considered as
rigorously localized. The size of the lowest magnitude eigenvalue can be used to characterize the 
extent of localization.

Power losses were calculated 
using the optical theorem. These are integrals over the cubic sample surface.
For the power absorbed we have 	          
\begin{equation}
P_{abs}=-\frac{c}{4\pi}Re\displaystyle{\oint}_{Surface}(\vec{E}\times\vec{H}^*)\cdot \vec{dS} 
\end{equation}
and the power scattered 
\begin{equation}
P_{scatt}=+\frac{c}{4\pi}Re\displaystyle{\oint}_{Surface}(\vec{E}_1\times\vec{H}_1^*)\cdot \vec{dS}  
\end{equation}
where $\vec{E}_1=\vec{E}-\vec{E}_o$ and $\vec{H}_1=\vec{H}-\vec{H}_o$ are the fields with the 
incident light fields subtracted out and Re means the real part.
The total losses are thus
\begin{equation}
P_{total} = P_{abs}+P_{scatt}. 
\end{equation}
Using Gauss' theorem and Maxwell's equations, these can be transformed into                    
\begin{equation}
 P_{total}= \frac{c k}{8\pi}Re\displaystyle{\int}_{vol.scatt.}i\Delta\epsilon^*\vec{E}^*\cdot\vec{E}_odv
\end{equation}
and
\begin{equation}
 P_{abs}= \frac{c k}{8\pi}Re\displaystyle{\int}_{vol.scatt.}i\Delta\epsilon^*\mid\vec{E}\mid^2dv.
\end{equation}
These are integrals over the scattering volume, which in this case is a sum over the volumes of the N scattering grains.
$P_{total}$ is of the same form as the extinction coefficent\cite{Schatz}.
Associated with each of the power losses there are path lengths, $l_{abs}$,$l_{scatt}$ and $l_{total}$ which are
calculated generically as
\begin{equation}
l = \frac{c v_{g}}{8\pi P}
\label{len}
\end{equation}
This is essentially the reciprocal of, the crossection associated with $P$ times the scatterer density,
where we assume one species and one grain size for simplicity.
For example, if we neglect multiple scattering and use the familiar
\begin{equation}
P^{dipole}_{scatt} = \frac{N k^{4}_{o} c \mid p_{o} \mid ^{2}}{3}
\end{equation}
where
\begin{equation}
p_{o} = \frac{\Delta\epsilon}{4 \pi}
\end{equation}
for unit incident field amplitude,
we can obtain a length $l^{dipole}_{scatt}$.
This power flow and length leave out multiple scattering effects.
   
Field energy densities were calculated using
\begin{equation}
w_E(\vec{r}) = Re\frac{\vec{E}(\vec{r})\cdot\vec{D}^*(\vec{r})}{16\pi},
\end{equation}
for the energy density in the electric field, and for the magnetic field energy density
\begin{equation}
w_H(\vec{r}) = Re\frac{\vec{H}(\vec{r})\cdot\vec{H}^*(\vec{r})}{16\pi} 
\end{equation}
where $\vec{D} = \epsilon\vec{E}$, for grid sites occupied by dielectric scattering grains, Re ${\epsilon} > 1$.
For metals, this is not the correct formulation of the grain energy as Re ${\epsilon} < 0$ and so 
this contribution is left out which
should make no difference for the comparison purposes needed here.
The total energy contained within the cubic sample is obtained by integrating these densities
over the sample cube
\begin{equation}
W_{total} = \displaystyle{\int}_{vol.sample}w\left(\vec{r}\right)d^3r.
\end{equation}
where $w\left(\vec{r}\right) = w_E\left(\vec{r}\right) + w_H\left(\vec{r}\right)$.
This total does not contain the far fields and can only be used as a relative measure
of energy concentration. We also calculate a figure of merit in analogy with resonant
circuits
\begin{equation}
Q(\omega) = \omega \left (\frac{W_{total}}{P_{total}}\right )
       \equiv  \omega \left (\frac{Field\hskip 2mm energy}  
         {Total\hskip 2mm Losses}\right ).  
\end{equation}

In order to characterize the energy density distribution, we calculated the energy density correlation function, defined as
\begin{widetext}
\begin{equation}
C\left(R\right) = \frac{d}{dR}\int_{vol.sample}d^{3}r\int_{0}^{|\vec{r}-\vec{r}^{\hskip .5mm '}|\leq R}
d^{3}r^{\hskip .5mm '}w\left(\vec{r}\right)w(\vec{r}+\vec{r}^{\hskip .5mm '}).
\label{Corr}
\end{equation}
\end{widetext}
The integrals were evaluated as sums over the sample lattice
points, the result smoothed and the R derivative taken by using a Savitzky-Golay 4 point filter\cite{NumRec}. In this calculation, the particle dimensions are equal to or smaller then the grid spacing
so the contributions to the integrals will ignore correlation in the same particle. The constant 
obtained from one particle will be obscured when the results normalized by their maximum value.
Also, the contribution to the sums is limited by the extent of the lattice grid; nevertheless,
Eq.(\ref{Corr}) gives some measure of the extent and position of energy localization.

\section{Model and Calculational Details}

\begin{figure}
\includegraphics[width=3.3in,angle=0]{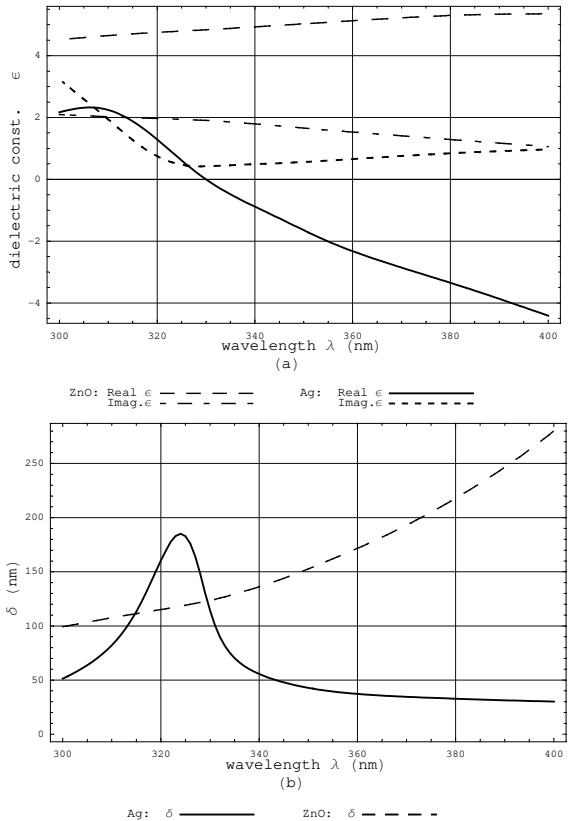}
\caption{\label{fig:dielectric} (a) The dielectric constants of ZnO(real part: long-dashed, imaginary part: dot-dashed) 
 and Ag(real part: solid, imaginary part: dashed) vs $\lambda$, for $300nm \leq \lambda \leq 400nm$. (b) The classical
 penetration depth for ZnO(long-dashed) and Ag(solid) in nm vs $\lambda$, for $300nm \leq \lambda \leq 400nm$.}
\end{figure}

The model used in generating most of our results consists of a cubic sample 1.6 microns on a side
with grid points inside this
sample. Using a grid, though not strictly necessary, enables the fields to be evaluated outside of the grains more easily. Having 32 grid divisions per side produces $32^3 = 32768$ sites in the sample. This means that the grid
interval is 50 nm. On a random set of grid points, we place our small spherical scattering grains with diameters
consistent with the long wavelength limit. We shall examine grains of ZnO with 50nm grain diameters and
grains of Ag with 20nm grain diameters. Figure~\ref{fig:50nmZnO900} gives an example of the sample with N = 900, 50nm diameter ZnO spherical grains. When considering light scattering, take the incident beam of light
to be a plane wave of fixed (unit) amplitude and phase going in the $\hat{z}$ direction with $\vec{E}$ along $\hat{x}$.
The range of wavelength considered is $300nm \leq \lambda \leq 400nm.$ Note that the plasma resonance of bulk Ag is at 
about 324 nm and that the excitonic emission wavelength that leads to lasing in semiconducting wurtzite structure ZnO corresponds to 385nm,
which is also close to the band gap for this material.
The ZnO scattering grains have a homogeneous dielectric constant whose frequency dependence is shown in Fig.~\ref{fig:dielectric}(a)
\cite{Mi}. Both the real and imaginary parts are given. Also shown is similar data for Ag \cite{PriSchatz}. 
Figure~\ref{fig:dielectric}(b) gives the calculated ``penetration depth'' corresponding to these dielectric constants\cite{Jackson}.

To find the lowest eigenvalues of the scattering matrix $\cal{P}$, the scattering problem was solved using
the conjugate gradient algorithm\cite{NumRec} with the fiduciary incident field described above. No transforms were done as
there is no benefit to be gained from either the fast Fourier or wavelet transforms when the scattering
centers are randomly distributed and there are no discernable internal boundaries. This may not be true, however,
for a very large number of scattering centers, $N \gg 900$, whereas $N = 900$ is as large a number as could
be treated reliably in the present calculations.

\section{Eigenvalue and Scattering Results}

\begin{figure}
\includegraphics[width=3.3in,angle=0]{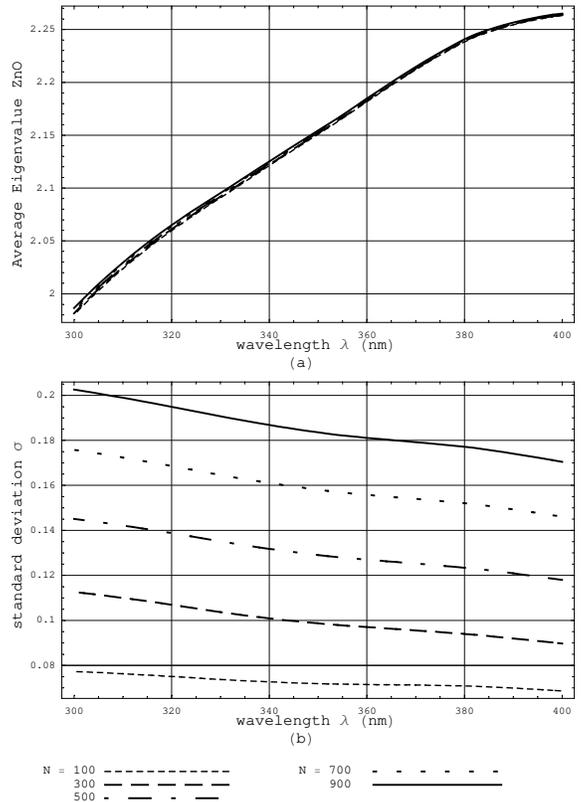}
\caption{\label{fig:ZnOaveig} (a) The magnitude of the average eigenvalue and the standard deviation ($\sigma$) of the eigenvalue magnitude
distribution vs $\lambda$ for 50nm diameter ZnO scattering grains. The average is approximately the same
for any number of scatterers. (b) The standard deviations are given for the number of scattering grains
in the sample: N = 100, 300, 500, 700 and 900 (see legend). }
\end{figure}

For the model described above, using 50nm diameter ZnO spherical scattering grains, the magnitudes of the eigenvalues of
the scattering operator given by Eq.(\ref{bigPolar}) were calculated. The results are summarized in Fig.(~\ref{fig:ZnOaveig}).
The average eigenvalue magnitude is the same for any number of scattering grains but the spread estimated by the standard deviation,
$\sigma$, increases with scattering number and thus the tail of the distribution extends more toward zero. There is a slight
shift of the average to higher values as the wavelength increases.
 
%
\begin{figure}
\includegraphics[width=3.3in,angle=0]{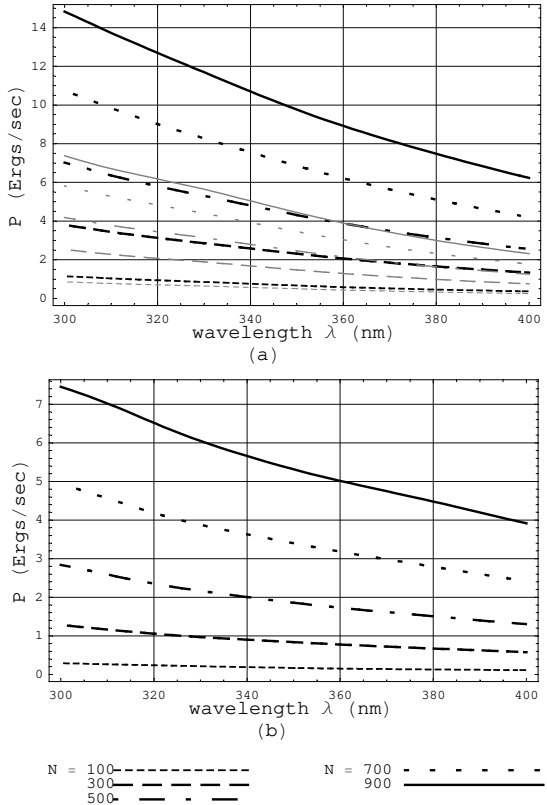}
\caption{\label{fig:ZnOloss} The losses for 50nm diameter ZnO scattering grains for a unit amplitude incident plane wave impinging
on the model sample Fig.(~\ref{fig:50nmZnO900}) from the negative z direction vs $\lambda$: (a) gives the total losses
$P_{total}$ (black heavy curves)
and the absorption losses $P_{abs}$ (gray lighter curves), the  number of scattering grains in the sample: N =   
100, 300, 500, 700 and 900 (see legend). (b) gives the scattering losses only $P_{scatt}$ for the corresponding
number of scatterers as in (a) above. }
\end{figure}
%
\begin{figure}
\includegraphics[width=3.3in,angle=0]{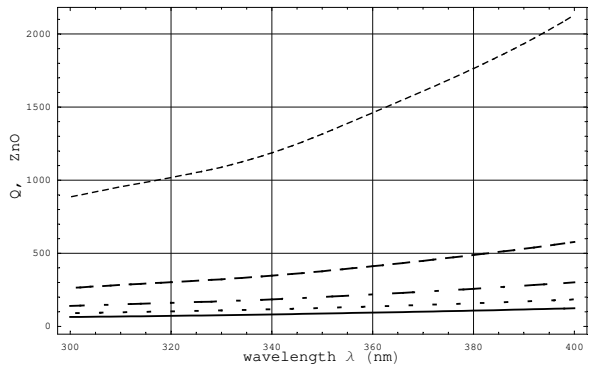} 
\caption{\label{fig:ZnOQ} The figure of merit Q defined in analogy with that calculated for resonant circuits for 50 nm diameter
ZnO scattering grains vs $\lambda$ and for the same range of scattering grain numbers considered above. }
\end{figure}
%
\begin{figure}
\includegraphics[width=3.3in,angle=0]{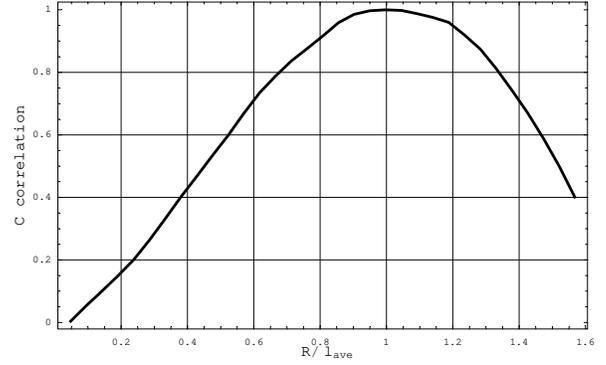}
\caption{\label{fig:ZCorr} The energy-energy correlation function C, for ZnO $\lambda = 385nm$ and $N = 900$. The y axis is normalized
to the maximum correlation  
and the x axis is normalized to the calculated average distance between scattering centers $l_{ave}$. This curve appears to be
universal for any number of scattering grains and at any wavelength; i.e. the curves for any wavelength, scattering
strength and particle number are practically identical and completely overlap.
The correlation function C, calculated for Ag also yields the same curve and
overlaps the one given above.}
\end{figure}
The scattering equation, Eq.(\ref{DisScatt}), was solved using an incident plane wave of unit amplitude impinging on the sample
volume from the z direction (see Fig.(~\ref{fig:50nmZnO900})). With the solution of the scattering problem, the losses and field
energies were calculated and the results for ZnO are summarized in Fig.(~\ref{fig:ZnOloss}). Part (a) of this figure gives
the total and absorptive losses and part (b) gives the scattering losses only. As expected, there is a decrease in these losses as the
wavelength increases. Also the losses due to scattering increase significantly as the particle number goes up.
In ZnO, the process happens in a very monotonic manner. These losses can also be analyzed in terms of lengths using Eq.(\ref{len}).
For example, for $N = 900$, $l_{total} \approx 0.10\lambda$ at $\lambda = 300(nm) \rightarrow l_{total} \approx 0.20\lambda$
at $\lambda = 400(nm)$ and
$l_{abs} \approx 0.22\lambda \rightarrow 0.55\lambda$, whereas $l_{scatt} \approx 0.15\lambda \rightarrow 0.21\lambda$.
The scattering length to wavelength ratios increase monotonically with incident wavelength as expected. Note that,
even for the largest number of scattering sites that can be used in this direct calculation, $l_{scatt}$ is still slightly
larger than the Ioffe-Regel value of $\approx .16\lambda$ over most of the wavelength range considered, and this is
with the average distance between scattering sites of 
$l_{av} \approx 2.9\lambda$ for $N = 900$. It seems, that, for realistic materials parameters, the Ioffe-Regel limit is
almost obtained in a direct calculation, however, no eigenvalues approach zero.
 
The relative figure of merit, or Q, for ZnO
is given in Fig.(~\ref{fig:ZnOQ}), and shows that Q decreases with an increase of scattering number. This means that losses
dominate and that there is no tendency to store energy. The  
energy-energy correlation function, which is defined in Eq.(\ref{Corr}), is given for ZnO in Fig.(~\ref{fig:ZCorr}). The solid lines
are for a wavelength of 385nm and $N = 900$. These results are normalized
to the maximum correlation value on the y axis and to the mean distance between scattering centers $l_{ave}$ on the x axis. There is
a broad maximum at $l_{ave}$, as expected if the energy is localized in and around the grains.
Note that, as explained above, there is no same particle correlation due to the small particle size
in relation to the grid spacing. There are no peaks off of this maximum
and the curve appears to be universal, i.e., it is practically the same for any number of scattering grains and at any wavelength
and for any scattering strength. This means that the curves for any values of these parameters 
completely overlap and are identical to Fig.(~\ref{fig:ZCorr}). It appears that this kind of energy-
energy correlation is quite general for point particle scattering. 
\begin{figure}
\includegraphics[width=3.3in,angle=0]{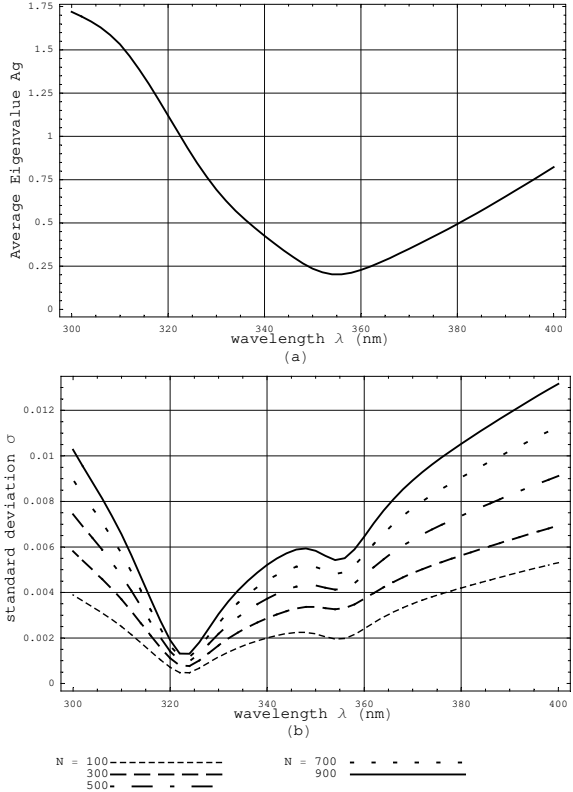}
\caption{\label{fig:Agaveig} (a) The magnitude of the average eigenvalue and the varience ($\sigma$) of the eigenvalue magnitude
distribution vs $\lambda$ for 20nm diameter Ag scattering grains. The average is the same
for any number scatterers. (b) The standard deviations are given by the dashed curves for the different
number of scattering grains in the sample: N =  100, 300, 500, 700 and 900 (see legend). }  
\end{figure}
%
\begin{figure}
\includegraphics[width=3.3in,angle=0]{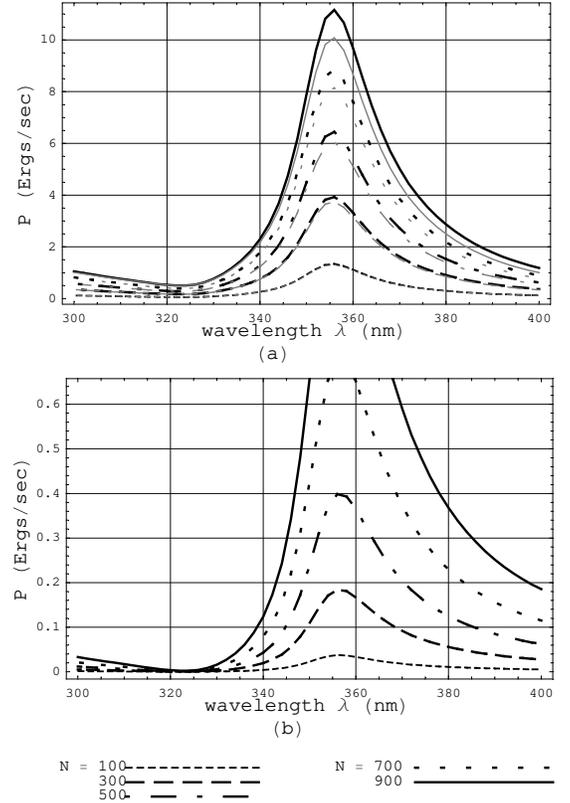}
\caption{\label{fig:Agloss} The losses for 20nm diameter Ag scattering grains for a unit amplitude incident plane wave impinging
on the model sample Fig.(~\ref{fig:50nmZnO900}) from the negative z direction vs $\lambda$: (a) the total losses $P_{total}$ (black heavy curves)
and the absorption losses $P_{abs}$ (gray lighter curves), the curves corresponding to the number of scattering grains in the sample: N =   
100, 300, 500, 700 and 900 (see legend). (b) the scattering losses only $P_{scatt}$ for the number of scatterers as in (a) above. }
\end{figure}
%
\begin{figure}
\includegraphics[width=3.3in,angle=0]{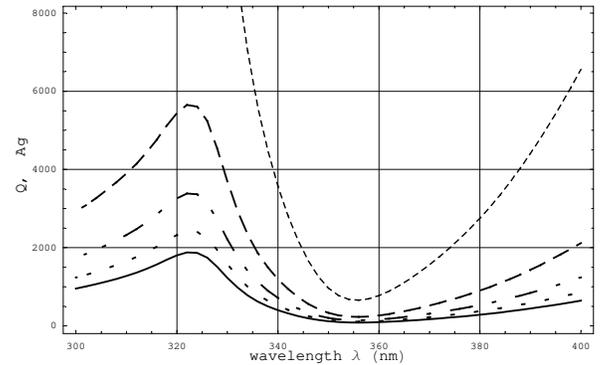} 
\caption{\label{fig:AgQ} The figure of merit Q defined in analogy with that calculated for resonant circuits for 20 nm diameter
Ag scattering grains vs $\lambda$ and for the same range of scattering grain number considered above. }
\end{figure}

Results analogous to those for ZnO are given for 20nm diameter spherical grains of Ag in Figs.(~\ref{fig:Agaveig}),
(~\ref{fig:Agloss}) and (~\ref{fig:AgQ}). Fig.(~\ref{fig:Agaveig}) for Ag eigenvalues is to be compared to Fig.(~\ref{fig:ZnOaveig}) for
ZnO eigenvalues. These results can be correlated with the 
dielectric properties shown in Fig.(~\ref{fig:dielectric}). 
For, Ag the scattering strength is resonantly large when the real part of the dielectric constant is near -2 which
is the Lorentz resonance and this yields the lowest
eigenvalues and a dip in the standard deviation. The sharpest dip in the eigenvalues, however, occurs where the imaginary
part of the dielectric constant is smallest. These effects become more pronounced as the number of scatterers increases. By
contrast the ZnO results show no such structure and only monotonic trends.
The eigenvalues for Ag are lower than those for ZnO and the spread is much narrower.
These effects are dramatically illustrated in the scattering losses. Fig.(~\ref{fig:Agloss}) shows the results
for Ag and compares to Fig.(~\ref{fig:ZnOloss}) for ZnO. Again the results for ZnO show nothing remarkable whereas the
Ag results have a large resonant peak and fall to a shallow minimum where the imaginary part of the dielectric constant
in minimal. The length results mirror the structure of the power loss results in an inverse fashion, however,
$l_{scatt}/\lambda$ on a scale one hundred times larger then either $l_{total}/\lambda$ or $l_{abs}/\lambda$. 
Even at the Lorentz resonance $l_{scatt} \approx 1.0\lambda$, however,
at $\lambda = 324(nm)$, where the imaginary part of $\epsilon$ is least, $l_{scatt} \approx 600\lambda$.
It is evident that length reduction due to losses is significant and that absorbtion accounts for most these losses. 

The Q results for Ag are given in Fig.(~\ref{fig:AgQ}) 
and are again dominated
by the scattering losses, though it is larger than that of ZnO, shown in Fig.(~\ref{fig:ZnOQ}),
there doesn't appear to be any tendency to store energy
since it also decreases with increasing scattering number. 
The energy-energy correlation corresponds to the universal curve given in Fig.(~\ref{fig:ZCorr}).
This means that all the calculations for Ag with various grain numbers and wavelengths result in the same
correlation curve C, which completely overlaps the one given in Fig.(~\ref{fig:ZCorr}).
\begin{figure}
\includegraphics[width=3.3in,angle=0]{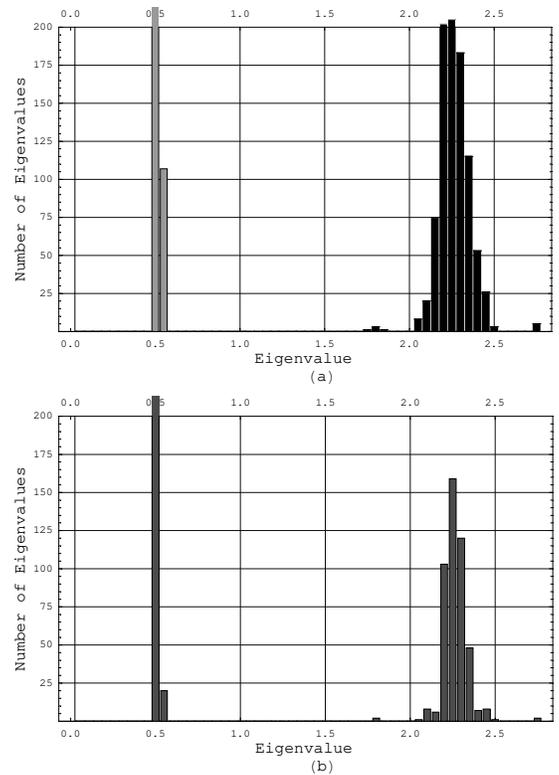} 
\caption{\label{fig:MixEigen} The eigenvalue spectrum of the scattering operator given as a histogram of the eigenvalue
magnitudes for $\lambda = 380nm$ and for: (a) 300 scattering grains of 50nm diameter ZnO in black and 300
scattering grains of 20nm diameter Ag in gray (b) a mixture of 150 grains of ZnO and 150 grains of Ag in dark gray. The eigenvalue range is
from 0 to 2.75 with a bin size of $\Delta = 0.05$. The y axis is truncated at a value of 200 for both (a) and (b). }
\end{figure}

The eigenvalue magnitude spectra for ZnO and Ag are widely separated, as shown in Fig.(~\ref{fig:MixEigen})(a) which gives results for 
300, 50nm diameter ZnO scattering grains and 20nm diameter Ag scattering grains at wavelength $\lambda = 380$nm 
graphed on the same scale.
The results run in value from zero to 2.75 with a  bin size of $\Delta = 0.05$. and the graph maximum
for the number of values in any bin is 200. The Ag results lie in a much narrower range and are lower in value than for ZnO.
When mixtures are considered 
Fig.(~\ref{fig:MixEigen})(b) is obtained. These results are due to a mixture of 150 ZnO grains and 150 Ag grains for the same
wavelength as used in Fig.(~\ref{fig:MixEigen})(a). One might expect that if there were any phase interference between light scattering
from these different species having widely differing scattering
strengths, that there would be some overlap of the eigenvalue spectra, but as seen in the figure, there is none.

\begin{figure}
\includegraphics[width=3.3in,angle=0]{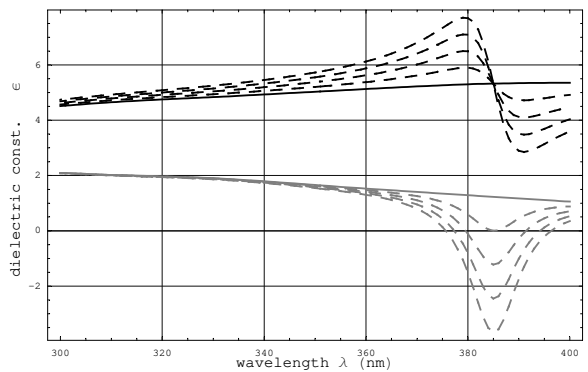} 
\caption{\label{fig:Gaindielec} The dielectric constant vs $\lambda$ for ZnO  where an additional resonant term has been added to
the backround constant for ZnO shown in Fig.(\ref{fig:dielectric}). The real part is in black and the imaginary part is in gray.
The resonance is located at $\lambda = 385nm$ with a width of 
$\gamma = 5*10^5 m^{-1}$ corresponding to the ZnO gap and pre-lasing lineshape given by Cao\cite{Cao}.
Gain levels of 0.0, -1.0, -2.0, -3.0, -4.0 x $10^{13}$ are shown
by the dashed lines. The imaginary part of the dielectric constant becomes more negative as the gain increases.  }
\end{figure}
%
\begin{figure}
\includegraphics[width=3.3in,angle=0]{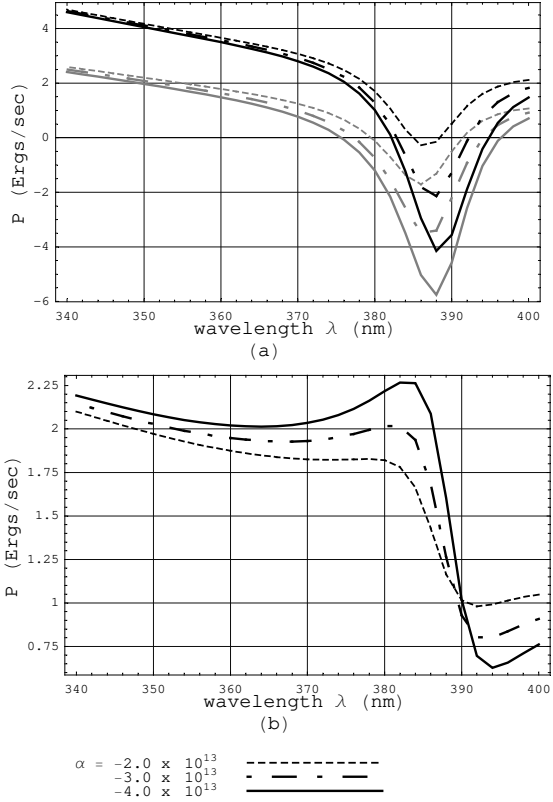} 
\caption{\label{fig:Gainloss} The losses for 500 ZnO scattering grains, 50nm in diameter and a unit amplitude incident plane wave impinging
on the sample from the negative z direction vs $\lambda$: (a) the total losses $P_{total}$ (black curves)
and the absorption losses $P_{abs}$ (gray curves) and (b) the scattering losses only $P_{scatt}$. 
The families of loss curves give results for
gains of -2.0, -3.0, -4.0 x $10^{13}$ with the curves dipping progressively more negative with increasing gain magnitude.}
\end{figure}
%
\begin{figure}
\includegraphics[width=3.3in,angle=0]{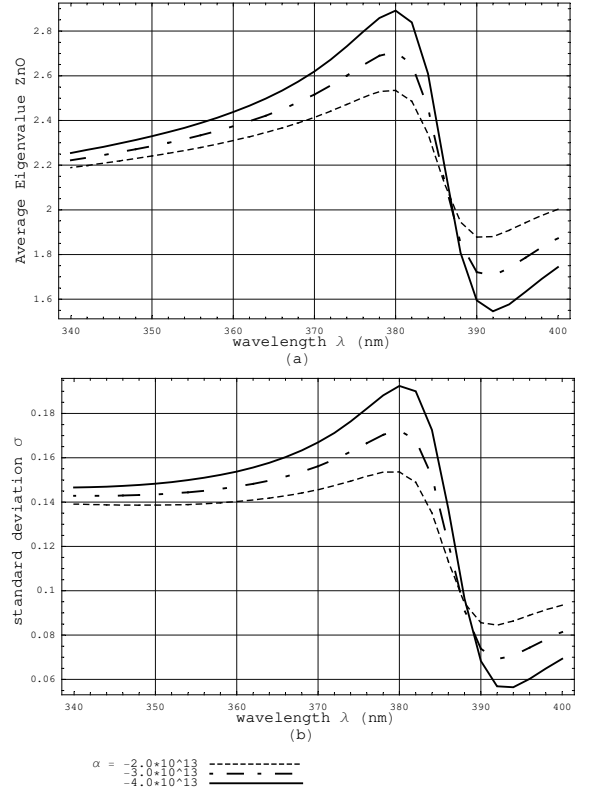} 
\caption{\label{fig:Gainaveig} (a) The magnitude of the average eigenvalue of the eigenvalue magnitude
distribution vs $\lambda$ for ZnO scattering sample given above for the corresponding gain levels. (b) The standard deviations ($\sigma$)
are given. We note that the distrbution of eigenvalue magnitudes changes somewhat; however, the lowest eigenvalue magnitude
does not move toward zero.  }
\end{figure}
\section{Effect of Adding Gain}

In order to explore the connection between light localization and lasing threshold, gain was added to
the ZnO system by including a resonant term to the background dielectric constant:
\begin{equation}
\epsilon\left(\vec{r}_{i}\right) = \epsilon_{B}\left(\vec{r}_{i}\right) + \frac{\alpha}{(k_{r}^{2}-k^{2}-i{\gamma}k)}.
\end{equation}
Here $\epsilon_{B}$ is the ZnO backround, $k_{r}$ is the resonant wavenumber taken to be $1.632*10^7 (1/m)$.
This corresponds to $\lambda = 385 nm$ which is at the bandgap in ZnO. The width $\gamma = 5*10^5 (1/m)$ is taken
from the pre-lasing lineshape given by Cao et al.\cite{Cao}. The gain is added by making ${\alpha}$ negative, simulating a
population inversion. Gain levels ${\alpha} = 0.0, -1.0, -2.0, -3.0, -4.0$ all x $10^{13} (1/m)^2$ were used. The dielectric constant for
these values in the range of wavelengths $340nm \rightarrow 400nm$ is shown in Fig.(~\ref{fig:Gaindielec}).
As gain is added the real part shows enhanced scattering on the higher
energy side of the resonance and suppressed scattering on the lower energy side.
The imaginary part below the x-axis and becomes more negative as more gain is added. To show that losses
can be compensated for, scattering results for these gains are given in Fig.(~\ref{fig:Gainloss})(a) and (b). These indicate the losses for
an incident planewave impinging on the sample from the negative z direction, as previously discussed, for our model. In this case, however,
results for 500 ZnO scattering grains are given but now with the indicated gains included in the dielectric constant. Note that 
total losses can be driven negative. Scattering losses are also affected, but of course remain positive. The only length, that
is non-singular and makes sense for this system, is $l_{scatt}$ which, mirror the $P_{scatt}$ results in an inverse fashion. 
For the largest gain, $l_{scatt} \approx 0.55\lambda$ at $\lambda = 383(nm)$, its lowest value, below the resonance and
$l_{scatt} \approx 2.0\lambda$ at $\lambda = 394(nm)$, its highest value, above it. 
Results are shown for
gains  $-2.0, -3.0, -4.0  x 10^{13} (1/m)^2$. Finally, Fig.(~\ref{fig:Gainaveig}) gives (a) the average eigenvalue magnitude and (b)
the standard deviations in the relevant frequency range for these gain levels. The spectrum is affected by the gain, the 
average dipping lower on the low energy side of the resonance, even though the scattering would seem to be somewhat suppressed 
there as indicated by the dielectric constant results. In any case, none of the eigenvalue magnitudes were driven toward zero.
This indicates that, at least in 3D systems, localization and lasing threshold are not connected in any direct manner and that
the interpretation of random laser experiments should be re-examined.  

\section{Conclusions}

It can be seen from the above results that there is very little energy concentrated between scattering particles, but that most 
of the energy is localized around and in the grains. Energy concentration due to phase interference does not seem to happen
for light in a 3D random scattering medium, at least for the system dimensions, number of scattering centers
and material parameters considered here.
There is no zero eigenvalue representing a local state and no isolated resonance near
zero that appears for a critical number of scattering particles. The Ioffe-Regel condition is not quite met with
realistic material parameters in this direct calculation.
However, the tendency to localization with randomness or with an increase
in the number and/or strength of the scatterers seeems to be gradual and continuous with no indication of any sharp transition. The energy-energy
correlation calculations, which show a broad peak at the mean particle separation, and the fact that the curve seems to be universal,
buttress this conclusion. Also, there is little interaction
between the spectra of scatterers of different strengths, as is indicated in the spectra of Ag and ZnO mixtures.

Adding gain to 
such a system of random scatterers, even enough to completely compensate for losses, does not cause the magnitude of any of the 
eigenstates to move to zero. This is in spite of the Ioffe-Regel condition not quite being met.
The lasing threshold can be considered as the point where the gain just compensates the losses.
Thus it would seem that localization and lasing threshold have little connection. This means that the 
interpretation of the mechanisms of random lasing should be revisited. While we have used 
an order of magnitude more scattering sites that was done previously,
it is clear that a still larger number of scattering sites would be of interest in order
to check for localization in realistic 3D random system. Experimental occupied volume fractions of 50\% are quoted\cite{Cao} and,
in the terms used here, that amounts to about $N \approx 30000$.
For this simulation, however, very different numerical techniques will
have to be used than have hitherto been applied.

\begin{acknowledgments}
The author gratefully acknowleges discussions with George Schatz, Hui Cao, A. J. Freeman and, especially, John Ketterson.
He is also grateful for the hospitality and support shown him by A. J. Freeman and the Condensed Matter/ Electronic Structure Theory Group at
Northwestern University. 
\end{acknowledgments}

\appendix
\section{Derivation of Green's Function}

Using the retarded potentials:
\begin{equation}
\vec{A}(\vec{r},t)=\frac{1}{c}\int\frac{\vec{j}(\vec{r}^{\hskip 1mm '},t^{\hskip .5mm '})d^3r^{\hskip .5mm '}}{R}
\label{A}
\end{equation}
and
\begin{equation}
\phi(\vec{r},t)=\int\frac{\rho(\vec{r}^{\hskip 1mm '},t^{\hskip .5mm '})d^3r^{\hskip .5mm '}}{R}
\label{phi}
\end{equation}
where $\rho$ and $\vec{j}$ are the source charge and current densities,
 $R=|\vec{r}-\vec{r}^{\hskip 1mm '}|$, and 
 $t^{\hskip .5mm '} = (t-\frac{R}{c}$) is the retarded time (i.e. the time that it takes light to travel from $\vec{r}^{\hskip 1mm '}$
to $\vec{r}$), the Lorentz gauge is chosen so that
\begin{equation}
\vec{\nabla}\cdot \vec{A}+\frac{1}{c}\frac{\partial\phi}
{\partial t} = 0.
\end{equation} 
The continuity equation 
for the source terms must be satisfied, i.e.
\begin{equation}
\vec{\nabla}^{\hskip .5mm '}\cdot \vec{j}+\frac{1}{c}\frac{\partial\rho}
{\partial t^{\hskip .5mm '}} = 0,
\end{equation} 
and this is manifestly true if the source terms are given by
\begin{equation}
\vec{j} = \frac{\partial\vec{P}}{\partial t^{\hskip .5mm '}} +
 c\vec{\nabla}^{\hskip .5mm '}
\times\vec{M} - \vec{\nabla}^{\hskip .5mm '}\cdot
\frac{\partial\stackrel{{\leftrightarrow}}{Q}}{\partial t^{\hskip .5mm '}} + ... 
\end{equation}
and
\begin{equation}
\rho = -\vec{\nabla}^{\hskip .5mm '}\cdot\vec{P} + \vec{\nabla}^{\hskip .5mm '}
\vec{\nabla}^{\hskip .5mm '}
:\stackrel{
\leftrightarrow}{Q} + ...
\end{equation}
which is also consistent with the Lorentz condition and the representation of the
potentials given in Eqs. (\ref{A}, \ref{phi}). Here $\vec{P}$ is the electric polarization
density, 
$\vec{M}$ the magnetic polarization
density, $\stackrel{\leftrightarrow}{Q}$ the electric quadrupole density, etc.
The derivatives here are taken with respect to the source location 
$\vec{r}^{\hskip 1mm '}$ and time $t'$. The elementary charges can
be grouped at a point into moments of any order to represent the material medium
both statically and dynamically; 
however, here we shall consider only electric dipole terms for simplicity and to illustrate 
the technique.

We now make explicit the derivatives with respect to the observer position contained in
the retarded time and write
\begin{widetext}
\begin{equation}
\vec{\nabla}^{\hskip .5mm '}\cdot[\vec{P}] = [\vec{\nabla}^{\hskip .5mm '}\cdot
\vec{P}] + \left[\frac{\partial\vec{P}}{\partial t}\right]\cdot\vec{\nabla}^{\hskip .5mm '}
\left(t - \frac{R}{c}\right) = [\vec{\nabla}^{\hskip .5mm '}\cdot\vec{P}] + 
\left[\frac{\partial\vec{P}}{\partial t}\right]\cdot\frac{\vec{R}}{cR}
\end{equation}
\end{widetext}
where $\vec{R} = \vec{r}-\vec{r}^{\hskip 1mm '}$ and [ ] denotes evaluation at retarded time, thus
\begin{widetext}
\begin{equation}
\phi(\vec{r},t) = -\int\frac{\left[\vec{\nabla}^{\hskip .5mm '}\cdot\vec{P}\right]}{R}
d^3r^{\hskip .5mm '} =
-\int\frac{\vec{\nabla}^{\hskip .5mm '}\cdot\left[\vec{P}\right]}{R} 
d^3r^{\hskip .5mm '}
+\int\frac{\left[\stackrel{\bullet}{\vec{P}}\right]\cdot\vec{R}}
{cR^2}d^3r^{\hskip .5mm '}.
\end{equation}
\end{widetext}
Also note that
\begin{equation}
\vec{\nabla}^{\hskip .5mm '}\cdot\left(\frac{\left[\vec{P}\right]}{R}\right) = \frac{\vec{\nabla}^{\hskip .5mm '}
\cdot\left[\vec{P}\right]}{R} + \left[\vec{P}\right]\cdot\vec{\nabla}^{\hskip .5mm '}
\left(\frac{1}{R}\right)
\end{equation}
so that
\begin{equation}
\vec{\nabla}^{\hskip .5mm '}\cdot\left(\frac{\left[\vec{P}\right]}{R}\right)
 - \frac{\left[\vec{P}\right]\cdot
\vec{R}}{R^3} =\frac{\vec{\nabla}^{\hskip .5mm '}\cdot\left[\vec{P}\right]}{R}.
\end{equation}
Therefore, we finally have:
\begin{equation}
\phi(\vec{r},t) = \int\frac{\left[\vec{P}\right]\cdot\vec{R}}{R^3}d^3r^{\hskip .5mm '} +
\int\frac{\left[\stackrel{\bullet}{\vec{P}}\right]\cdot\vec{R}}{cR^2}d^3r^{\hskip .5mm '}
\label{exp_phi}
\end{equation}
where $\stackrel{\bullet}{\vec{P}}$ indicates a derivative with respect to time.
In a similar way we obtain
\begin{equation}
\vec{A}(\vec{r},t) = \frac{1}{c}\int\frac{\left[\stackrel{\bullet}{\vec{P}}\right]}{R}
d^3r^{\hskip .5mm '}.
\label{exp_A}
\end{equation}

The fields are given by:
\begin{equation}
\vec{E} = -\frac{1}{c}\frac{\partial\vec{A}}{\partial t} - \vec{\nabla}\phi
\end{equation}
and
\begin{equation}
\vec{H} = \vec{\nabla}\times\vec{A}. 
\end{equation}
We need only solve the equation for the $\vec{E}$  field because from it we can obtain $\vec{H}$.

Up to this point, we have made no assumption about the time dependence but in order to 
make contact with the Green's function formulation of the scattering problem
\cite{Martin} and with the coupled dipole approach\cite{Schatz} we shall assume a monochromatic source
$\sim e^{-i\omega t}$. With this assumption:
\begin{equation}
\stackrel{\bullet}{\vec{P}}\left(t\right) = -i\omega\vec{P}\left(\omega\right)e^{-i\omega t}
\end{equation}
\begin{equation}
\left[\stackrel{\bullet}{\vec{P}}\left(t\right)\right] = -i\omega\vec{P}\left(\omega\right)
e^{-i\omega\left(t - \frac{R}{c}\right)}.
\end{equation}
Letting $k = \frac{\omega}{c}$ we have
\begin{equation}
\left[\stackrel{\bullet}{\vec{P}}\left(t\right)\right] = -ick\vec{P}
e^{i\left(kR-\omega t\right)}.
\end{equation}
Note that the 
important $e^{ikR}$ phase term is due to evaluation at the retarded time.
The potentials are now given as
\begin{equation}
\vec{A}(\vec{r},t) = -ik\int d^3r^{\hskip .5mm '}\frac{\vec{P}\left(\vec{r}^{\hskip .5mm '}
\right)}{R} e^{i\left(kR-\omega t\right)},
\end{equation}
so
\begin{equation}
-\frac{1}{c}\frac{\partial \vec{A}}{\partial t} = k^2\int d^{3}r^{\hskip .5mm '}
\frac{\vec{P}\left(\vec{r}^{\hskip .5mm '}
\right)}{R} e^{i\left(kR-\omega t\right)},
\end{equation}
and
\begin{equation}
\phi (\vec{r},t) = \int d^{3}r^{\hskip .5mm '}\vec{R}\cdot\vec{P}\left(\vec{r}^{\hskip .5mm '}\right)
\left(\frac{1}{R^{3}} - \frac{ik}{R^{2}}\right) e^{i\left(kR -\omega t\right)}.
\end{equation}
Using Eq.(\ref{exp_A}) and Eq.(\ref{exp_phi}). Factoring out the $e^{-i\omega t}$ time dependence,
we have for the $\vec{E}$ field:
\begin{widetext}
\begin{equation}
\vec{E}\left(\vec{r}\right) = \vec{E}_{o}\left(\vec{r}\right) +
k^{2}\int d^3r^{\hskip .5mm '}\frac{\vec{P}\left(\vec{r}^{\hskip .5mm '}\right)}
{R}e^{ikR} + \vec{\nabla}\int d^{3}r^{\hskip .5mm '}\left(\frac{ik}{R} - \frac{1}{R^{2}}
\right)\frac{\vec{R}\cdot\vec{P}\left(\vec{r}^{\hskip .5mm '}\right)}{R}
e^{ikR}
\label{Efield}
\end{equation}
\end{widetext}
where we have added an external "incident" field which comes from other sources not
given by the potentials considered above; this is the inhomogeneous term in the light
scattering equation. The $\vec{H}$ field can be obtained from
\begin{equation}
\vec{H}(\vec{r}) = -\frac{i}{k}\vec{\nabla}\times\vec{E}
\label{Hfield}
\end{equation}
which is Faraday's law using our monochromatic assumption.
Note that the $\vec{\nabla}$ term drops out when the curl in the above equation is applied
to the $\vec{E}$ field given in Eq.(\ref{Efield}).
The scalar retarded Green's function is defined as
\begin{equation}
g\left(R\right) = \frac{e^{ikR}}{4\pi R}
\end{equation}
so that
\begin{equation}
\vec{\nabla}g\left(R\right) = g\left(R\right)\left(\frac{ik}{R} - \frac{1}{R^{2}}\right)\vec{R}.
\end{equation}
Using this relation Eq.(\ref{Efield}) becomes Eq.(\ref{Scatt}) in the main text above.
%
%

%

\begin{thebibliography}{99}
\bibitem{Lawandy}
N. M. Lawandy et al, Nature (London) 368, 436 (1994)
\bibitem{Letokhov}
V. S. Letokhov,Sov. Phys. JETP 26, 835 (1968)
\bibitem{Wiersma}
D. S. Wiersma et al, Nature (London) 390, 671 (1997)
\bibitem{ChabStoy}
A. A  Chabanov, M. Stoytchev and A. Z. Genack,
Nature (London) 404, 850 (2000)
\bibitem{Anderson}
P. W. Anderson, Phys. Rev. 109, 1492 (1958)
\bibitem{John}
Sajeev John, Physics Today, pg. 32 (May 1991) and references therein.
\bibitem{Ioffe}
A. F. Ioffe and A. R. Regel, Prog. Semicond. 4, 237-291 (1960)
\bibitem{Chabanov}
A. A. Chabanov and A. Z. Genack,
Phys. Rev. Letters 87, 15, 153901 (2001)
\bibitem{Poles}
M. Rusek, A. Ortiowski and J. Mostowski,
Phys Rev. E 53, 4, 4122 (1996)
\bibitem{CaoXu} 
H. Cao, J. Y. Xu, S. H. Chang and S. T. Ho,
Phys. Rev. E 61, 2, 1985 (2000)
\bibitem{Cao}
H. Cao et al,
Phys. Rev. Letters 82, 11, 2278 (1999) and
Phys. Rev. Letters 84,  24,5584 (2000)
\bibitem{Burin}
A. L. Burin et al,
Phys. Rev. Letters 87, 21 (2001)
\bibitem{Qiming}
Qiming Li, K. M. Ho and C. M. Soukoulis
Physica B 296, 78 (2001)
\bibitem{Soest}
Gijs van Soest, F. J. Poelwijk, R. Sprik and Ad Lagendijk,
Phys. Rev. Letters 86,8, 1522 (2001)
\bibitem{Vanneste}
C. Vanneste and P. Sebbah,
Phys. Rev. Letters 87, 18, 183903-1 (2001)
\bibitem{Martin}
O. J. F. Martin and N. B. Piller,
Phys. Rev. E 58, 3, 3909 (1998)
\bibitem{Periodic}
R. P. Kenen and P. R. Sievert,      
J. Comp. Physics 14, 350 (1974)
\bibitem{Chew}
W. C. Chew;,
Waves and Fields in Inhomogeneous Media, Chapter 7;
Van Nostrand Reinhold, (1990)
\bibitem{Yaghjian}
A. D. Yaghjian;
Poceedings of the IEEE, 68, (1980).
\bibitem{Born}
M. Born and E. Wolf;
Principles of Optics, 5th edition, Chapter 2.4.1 and Appendix V;
Pergamon  Press, (1975).
\bibitem{Lorentz}
H. A. Lorentz, The Theory of Electrons, pg.138, 2nd Ed., Dover, New York (1952)
\bibitem{Meir}
M. Meier and A. Wokaun;
Optics Letters  Vol. 8, No. 11, November 1983.
\bibitem{Wokaun}
A. Wokaun, J. P. Gordon, and P. F. Liao,
Phys. Rev. Letters 48, Number 14, (1982)  
\bibitem{Schatz}
T. Jensen, L. Kelly, A. Lazarides and G. Schatz;
J. Cluster Science 10, No 2, (1999)
\bibitem{Druger}
S. D. Druger and B. V. Bronk;
J. Opt. Soc. Am. B Vol. 16, No. 12, 2239 (1999).
\bibitem{Mi}
Y. Mi, H. Odaka, and S. Iwata;
Japanese. J. Applied Phys. 38, 6A, 3453 (1999).
\bibitem{PriSchatz}
G. Schatz, Northwestern University, private communication.
\bibitem{Jackson}
W. D. Jackson;
Classical Electrodynamics 2nd Ed.,
John Wiley \& Sons, Inc.
\bibitem{NumRec}
W. H. Press, S. A. Teukolsky, W. T. Vetterling, and B. P. Flannery;
Numerical Recipes in C, Second Edition,
Cambridge University Press.
\end{thebibliography}
\end{document}